\documentclass[letterpaper,english,aps,prb,floatfix,twocolumn,amsfonts,superscriptaddress,longbibliography]{revtex4-2}

\usepackage[T1]{fontenc}
\usepackage[latin1]{inputenc}
\usepackage{graphics}
\usepackage{amssymb}
\usepackage{amsmath}
\usepackage{overpic}

\newcommand{\be}{\begin{equation}}
\newcommand{\ee}{\end{equation}}
\newcommand{\bea}{\begin{eqnarray}}
\newcommand{\eea}{\end{eqnarray}}

\def\g{\gamma}

\def\G{\Gamma}

\def\bk{{\bf k}}

\def\bq{{\bf q}}

\def\nn{\nonumber}
\def\lb{\label}
\def\pref#1{(\ref{#1})}

\newcount\bozza \bozza=0
\ifnum\bozza=1
\newdimen\shift \shift=-2truecm
\def\lb#1{%
{\label{#1}\rlap{\kern\shift{$\scriptstyle#1$}}}}
\else\def\lb#1{\label{#1}} \fi

\begin{document}
\title{Comment on: Calculation of an Enhanced $A_{1g}$ Symmetry Mode Induced by Higgs Oscillations in the Raman Spectrum of High-Temperature Cuprate Superconductors}

\author{Lara Benfatto}
\affiliation{Department of Physics and ISC-CNR, ``Sapienza'' University of Rome, P.le A. Moro 5, 00185 Rome, Italy}
\author{Claudio Castellani}
\affiliation{Department of Physics and ISC-CNR, ``Sapienza'' University of Rome, P.le A. Moro 5, 00185 Rome, Italy}
\author{Tommaso Cea}
\affiliation{IMDEA Nanoscience, C/Faraday 9, 28049 Madrid (Spain)}
\date{\today}

\maketitle

In a recent Letter \cite{puviani_prl21} Puviani et al. claim that the Higgs mode gives an enhancement of the Raman response in the symmetric $A_{1g}$ Raman channel. Here we report technical mistakes which invalidate this conclusion.
The Raman susceptibility is the response function for the Raman density operator $\rho_R=\sum_\bk \gamma_\bk c^\dagger_{\bk}c_{\bk}$, where the combination of incident/scattered light enters in each Raman form factor $\gamma_\bk$ (Ref. \cite{deveraux_review}). 
The authors of Ref.\ \cite{puviani_prl21} consider a model $H=H_0+H_P+H_C$, where $H_0$ is the non-interacting Hamiltonian, $H_C$ is the Coulomb interaction and $H_P$ is the four-fermion interaction term, Eq. (1) of \cite{puviani_prl21}, that can be written in Nambu notations as:
\bea
H_P&=&\sum_{\bq,\bk,\bk'} V(\bk,\bk'\bq) \psi^\dagger_{\bk+\bq/2}\tau_1\psi_{\bk-\bq/2}\psi^\dagger_{\bk'-\bq/2}\tau_1\psi_{\bk'+\bq/2}+\nn\\
\lb{ham}
&+&\sum_{\bq,\bk,\bk'} V(\bk,\bk'\bq) \psi^\dagger_{\bk+\bq/2}\tau_2\psi_{\bk-\bq/2}\psi^\dagger_{\bk'-\bq/2}\tau_2\psi_{\bk'+\bq/2}. 
\eea
In Ref.\ \cite{puviani_prl21} the authors further assume a  separable potential $V(\bk,\bk'\bq)\simeq -(V/4)f_\bk f_{\bk'}$. Such an approximation is justified for Eq.\ \pref{ham} at momenta $\bq\simeq 0$, where Eq.\ \pref{ham} reduces to residual $d$-wave pairing interaction between fermions. The Raman response at BCS level accounts only for the effects of the superconducting (SC) gap opening, and are encoded in the bare Raman response function $\chi_{\g\g}$. To account for interaction effects beyond mean-field level, including the Higgs-mode fluctuations, one should compute a dressed susceptibility $\chi^{full}_{\g\g}$. As it is has been discussed in several textbooks\cite{schrieffer,mahan} and routinely applied to the case of Raman response\cite{deveraux_review,klein_prb84,maiti_prb17}, in the dressed susceptibility {\em one} bare Raman vertex $\gamma_\bk$ is replaced by the dressed vertex $\Gamma_{\bk,m}$, which accounts for 4-fermion interactions included in $H_C$ and $H_P$. In the same notations of Ref.\ \cite{puviani_prl21}, such $\chi^{full}_{\g\g}$ is:
\bea
\lb{chiren}
\chi^{\text{full}}_{\gamma\gamma}(i\Omega_m)&=&
T\sum_{\bk,i\nu_n}\mathrm{Tr}\left[G_{\bk,n}\g_\bk\tau_3G_{\bk,n+m}\G_{\bk,m}\right]
\eea
This differs from  Eq.\ (7) of Ref.\ \cite{puviani_prl21}, where {\em both} vertices are renormalized, leading to overcounting of the diagrams and uncontrolled results for the Higgs-mode contribution. This is the first main observation, which invalidates the conclusions of Ref.\ \cite{puviani_prl21}. 

A second observation concerns the form of the vertex corrections, Eq.\ (5) in Ref.\ \cite{puviani_prl21}. Vertex corrections from the second line of Eq.\ \pref{ham} account for the effect of SC phase fluctuations on the Raman response. Along with Coulomb interactions, they are needed to restore gauge invariance of the Raman response in the limit where $\gamma_\bk$ is a constant\cite{deveraux_review,klein_prb84,cea_prb16,cea_leggett_prb16,maiti_prb17}. The result is encoded in the screened $\tilde \chi$ bubbles  of Eq. (4) of Ref.\ \cite{puviani_prl21}. Vertex corrections in the Higgs channel follow from the first term of Eq.\ \pref{ham}:
\be
\lb{ver1}
\Gamma_{\bk,m}=\tau_3\g_\bk-\frac{V}{2}\tau_1f_\bk T\sum_{\bk',n'} f_{\bk'}\mathrm{Tr}\left[\tau_1G_{\bk',n'+m}\Gamma_{\bk',m}G_{\bk',n'}\right].
\ee
This equation can be readily derived from the first line of Eq.\ \pref{ham} by building up the diagrammatic series with $H_P$ constrained to $\bq\simeq 0$, where a separable $d$-wave pairing interaction is justified. It differs, however, from Eq.\ (5) of Ref.\ \cite{puviani_prl21}, where the authors have included additional contributions from Eq.\ \pref{ham} at {\em large} $\bq$ values. In addition, even in this case the momentum dependence and the sign should differ from the one of Eq. (4) of Ref.\ \cite{puviani_prl21}, since it would read:
 \be
 \lb{ver2}
 \G_{\bk,m}=\gamma_\bk\tau_3+\frac{V}{2}T\sum_{\bk',i\nu_n'} f^2\left(\frac{\bk-\bk'}{2}\right)\left[\tau_1 G_{\bk',n'+m}\G_{\bk',m}G_{\bk',n'}\tau_1\right].
 \ee
Eq.\ \pref{ver2} contains corrections both in the particle-particle and in the particle-hole channels. The correct procedure to include into the problem particle-hole  interactions is to start from a general gauge-invariant interaction and then project it out in all channels. 
From the correct vertex equation in the Higgs channel, i.e. Eq.\ \pref{ver1}, one readily obtains the second term of Eq. (4) in Ref. \cite{puviani_prl21}, showing that this expression already accounts for Higgs fluctuations at the level of ladder resummation of the interaction \pref{ham}.  
This equation has been previously obtained in Ref.\ \cite{cea_prb16} for the $s$-wave case ($\gamma_\bk=1$), where it was shown that the Higgs contribution to the Raman response is quantitatively irrelevant, as also confirmed by the manuscript by Puviani et al (see line QP+CF+AM in Fig. 6 of Supplementary). On the other hand, the numerical results based on Eq. (7) of Ref.\ \cite{puviani_prl21} are derived from a wrong susceptibility because: (i) correcting two vertices leads to overcounting of diagrams and (ii) the vertex equation itself does not correspond to fluctuations in the Higgs channel. 

\bibliographystyle{apsrev}
\bibliography{pump-probe.bib}

\end{document}